# Pressure-Driven Metallicity in Ångström-Thickness 2D Bismuth and Layer-Selective Ohmic Contact to MoS$_2$

Shuhua Wang,[1] Shibo Fang,[1]* Qiang Li,[2] Yunliang Yue,[3] Zongmeng Yang,[4] Xiaotian Sun,[5] Jing Lu,[4] Chit Siong Lau,[6,7,1] L. K. Ang,[1] Lain-Jong Li,[8] and Yee Sin Ang[1]*

[1] Science, Mathematics and Technology (SMT) Cluster, Singapore University of Technology and Design, Singapore 487372

[2] Department of Physics, Hubei Minzu University, Enshi, 445000, P. R. China.

[3] School of Information Engineering, Yangzhou University, Yangzhou 225127, China

[4] State Key Laboratory for Mesoscopic Physics and School of Physics, Peking University, Beijing 100871, P. R. China

[5] College of Chemistry and Chemical Engineering, and Henan Key Laboratory of Function-Oriented Porous Materials, Luoyang Normal University, Luoyang 471934, P. R. China

[6] Quantum Innovation Centre (Q. InC), Agency for Science Technology and Research (A*STAR), Singapore 138634

[7] Institute of Materials Research and Engineering (IMRE), Agency for Science Technology and Research (A*STAR), Singapore 138634

[8] Department of Materials Science and Engineering, National University of Singapore 117575

*Corresponding Authors: shibo_fang@sutd.edu.sg, yeesin_ang@sutd.edu.sg

## Abstract

Recent fabrication of two-dimensional (2D) metallic bismuth (Bi) via van der Waals (vdW) squeezing method opens a new avenue to *ultrascaling* metallic materials into the ångström thickness regime [Nature 639, 354 (2025)]. However, freestanding 2D Bi typically exhibits a semiconducting phase [Nature 617, 67 (2023), Phys. Rev. Lett. 131, 236801 (2023)], which contradicts the experimentally observed metallicity in vdW-squeezed 2D Bi. Here we show that such discrepancy originates from the pressure-induced buckled-to-flat structural transition in 2D Bi, which changes the electronic structure from semiconducting to metallic phases. Based on the experimentally fabricated MoS$_2$-Bi-MoS$_2$ *trilayer* heterostructure, we demonstrate the concept of *layer-selective Ohmic contact* in which one MoS$_2$ layer forms an Ohmic contact to the sandwiched Bi monolayer while the opposite MoS$_2$ layer exhibits a Schottky barrier. The Ohmic contact can be switched between the two sandwiching MoS$_2$ monolayers by changing the polarity of an external gate field, thus enabling charge to be spatially injected into different MoS$_2$ layers. The layer-selective Ohmic contact proposed here represents a *layertronic* generalization of metal/semiconductor contact, paving the way towards layertronic device application.

Recent experimental synthesis of ultrathin 2D monolayer metals – including Bi, Ga, In, Sn, and Pb – using van der Waals (vdW) squeezing method with $MoS_2$[1] opens an avenue towards the synthesis of ultrathin ångström-thickness 2D metal. The ultrascaling of Bi into the ångström thickness limit is particularly attractive for 2D semiconductor device applications such as field-effect transistor (FET). [2–7] $MoS_2$, as one of the most extensively studied 2D materials, finds wide applications in various fields such as field-effect transistors,[8,9] valleytronics,[10,11] layertronics,[12–14] and spintronics [15,16] Bulk Bi forms ultralow contact resistance, Ohmic contact to $MoS_2$, [5,17–24] arising from the appropriate band alignment and the semimetallic nature of Bi [25–27]. As the detrimental Fermi level pinning (FLP) effect can be strongly suppressed in 2D vdW metal contact to 2D semiconductor,[28] 2D metallic Bi represents an exciting electrode material candidate to alleviate the FLP effect in 2D semiconductor FET.

However, vdW-squeezed 2D Bi[1] exhibits several inconsistencies with prior studies. Firstly, the vdW-squeezed 2D Bi exhibits metallic behavior as revealed by temperature-dependent transport measurements. However, first-principles calculations[29–31] and prior experimental studies of 2D Bi [32] typically observe a narrow-gap ferroelectric semiconducting behaviour with a bandgap of approximately 0.27 eV. Secondly, previous theoretical studies of 2D Bi typically adopt a buckled structure,[29,33,34] whereas transmission electron microscopy (TEM) images reveal an atomically flat layered structure in vdW-squeezed monolayer Bi.[1] Understanding the origin of metallicity and the structural properties of vdW-squeezed 2D Bi is thus a critical quest before their full potential can be unlocked.

Here, using first-principles density functional theory (DFT) calculations, we show that metallicity can arise from the out-of-plane pressure exerted on the 2D Bi during the vdW squeezing fabrication. Under elevated pressure, we show that 2D Bi undergoes a buckled-to-flat structural transition accompanied by a semiconducting-to-metallic phase transition. Such a phase transition corroborates well with the flat structure and the metallicity observed in vdW-squeezed 2D Bi. We further investigate the contact and transport properties between metallic flat 2D Bi and $MoS_2$ based on the experimentally fabricated $MoS_2$-Bi-$MoS_2$ *trilayer* heterostructure. [1] We propose the concept of *layer-selective Ohmic contact* in which one $MoS_2$ layer exhibits Ohmic contact to the sandwiched Bi monolayer while the opposite layer exhibits a Schottky barrier. Such layer-selective Ohmic contact can be spatially switched between the two sandwiching MoS$_2$ monolayers by changing the polarity of out-of-plane electric field of about 0.15 V/Å (or 15 MV/cm). The proposed concept of layer-selective Ohmic contact represents a *layertronic* generalization of metal/semiconductor (MS) contact, enabling gate-tunable charge injection to be layer selectively injected into different MoS$_2$ layers. Our findings shed light on the physical origin of metallicity in vdW-squeezed 2D Bi and unveils the potential of their contact heterostructure with MoS$_2$ for layertronic device application.

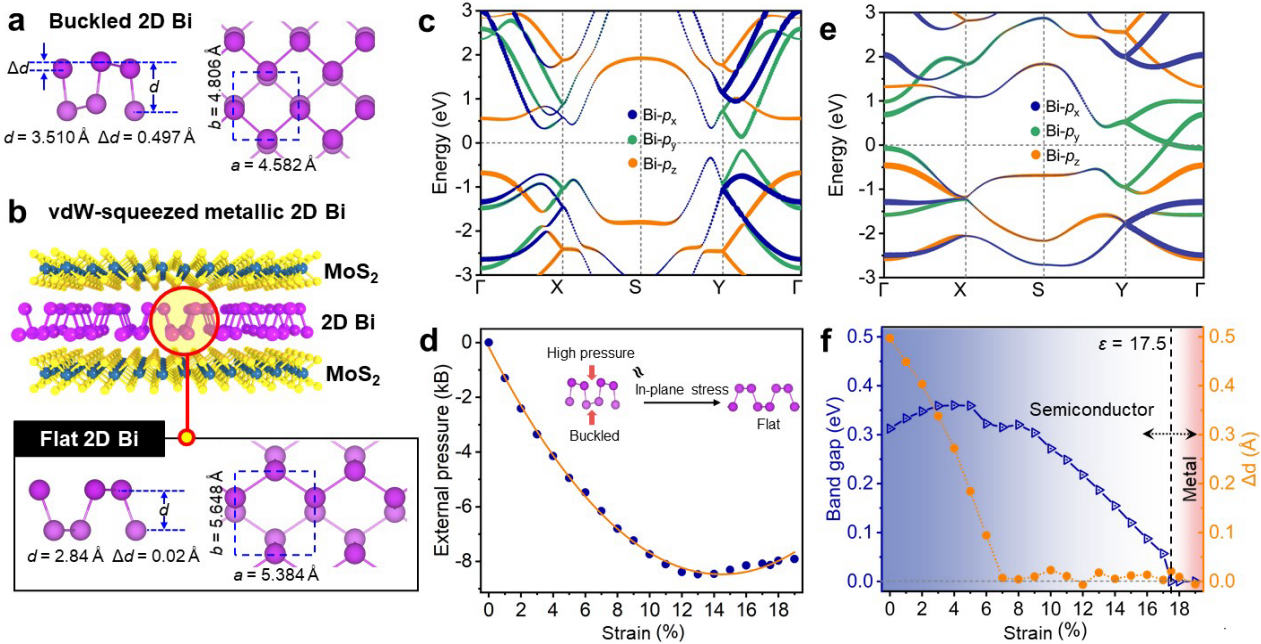

**Figure 1. Lattice and electronic structures evolutions of 2D Bi under external pressure.** Top and side views of 2D Bi in the (a) buckled (semiconducting) and (b) flat (metallic) configurations, with the layer thickness (*d*) and out-of-plane corrugation (Δ*d*) labeled. (c) Projected electronic band structures of 2D buckled Bi. (d) Schematic diagram of the structural deformation of 2D Bi under external pressure and the relationship between applied strain and pressure. (e) Same as (c) but for 2D flat Bi. (f) Evolution of the band gap (left axis) and out-of-plane corrugation Δ*d* (right axis) as a function of in-plane stress. A semiconductor-to-metal phase transition occurs at a critical strain of ε = 17.5%.

## Pressure-driven origin of metallicity in vdW-squeezed 2D metallic Bi

The atomic structure of 2D semiconducting Bi is shown in **Figure 1a**, with lattice constants of *a* = 4.582 Å and *b* = 4.806 Å, and belonging to the *Pmn*2$_1$ space group. This Bi monolayer exhibits a buckled geometry, as characterized by a layer thickness of 3.51 Å and an out-of-plane buckling of Δ*d* = 0.497 Å, which is defined as the vertical distance between the two outermost Bi atoms. Buckled 2D Bi exhibits physical properties such as negative Poisson's ratio, in-plane ferroelectricity, and negative piezoelectricity.[29,32,34,35] Their electronic band structure reveals a semiconducting nature with a bandgap of 0.31 eV [**Figure 1c**]. However, as revealed in the TEM images, the experimentally synthesized metallic 2D Bi exhibits a flat and layered structure without apparent buckling.[1] To understand this discrepancy, we applied an in-plane tensile strain to the semiconducting Bi monolayer corresponding to an out-of-plane compressive stress [**Figure 1d**], to mimic the experimental conditions under which 2D metallic Bi was synthesized under the vdW squeezing via $MoS_2$ layers. For each strained configuration, the out-of-plane (z-direction) stress component was extracted from the stress tensor computed by VASP.[36] In-plane tensile strain applied to an initially buckled Bi monolayer induces a contraction in the out-of-plane direction due to the Poisson effect, which leads to a progressively larger compressive stress along the z direction (i.e., more negative pressure). The pressure reaches its peak value near the 15% strain, and slightly increases at higher strains, indicating softening of the structure.

As shown in **Figure 1e,f**, when the tensile strain reaches 17.5%, a semiconductor-to-metal transition occurs due to the band gap closure (see Supplementary **Figure S1** for details), and the buckling is entirely flattened with $\Delta d$ = 0. Flat 2D Bi exhibits lattice constants of *a* = 5.384 Å and *b* = 5.648 Å under *Pmn*$2_1$ space group. As shown in the projected electronic structure calculations in **Figure 1e**, flat 2D Bi becomes a semimetal with conduction band minimum (CBM) touching the valence band maximum (VBM) at the Fermi level.

We next analyze the mechanism behind the semiconductor-to-metal transition of monolayer Bi under increasing in-plane tensile strain. As shown in **Figure 2a**, the buckled Bi structure leads to two inequivalent Bi atoms, labeled Bi1 and Bi2. The in-plane $p_{x/y}$ orbitals of Bi1 and Bi2 interact to form a filled bonding state (σ) and an unoccupied antibonding state (σ*), separated by a relatively large energy gap. The buckled structure gradually flattens after applying in-plane tensile strain, and the distance between Bi1 and Bi2 increases. This increased separation weakens the orbital hybridization between Bi1-$p_{x/y}$ and Bi2-$p_{x/y}$ orbitals, leading to a reduced bonding–antibonding splitting [**Figure 2b**]. As a result, both the bonding and antibonding states move closer toward the Fermi level. The projected band structures in **Figure 2c,d** show that the semiconductor-to-metal transition is primarily driven by the band gap reduction associated with $p_y$–$p_y$ orbital interactions. pCOHP analysis further confirms this evolution between Bi1-$p_y$ and Bi2-$p_y$. Relative to the buckled Bi structure [**Figure 2c**], the antibonding states (σ*) in flat Bi [**Figure 2d**] shift downward while the bonding states (σ) move upward and eventually cross the Fermi level, thereby giving rise to the emergence of metallicity.

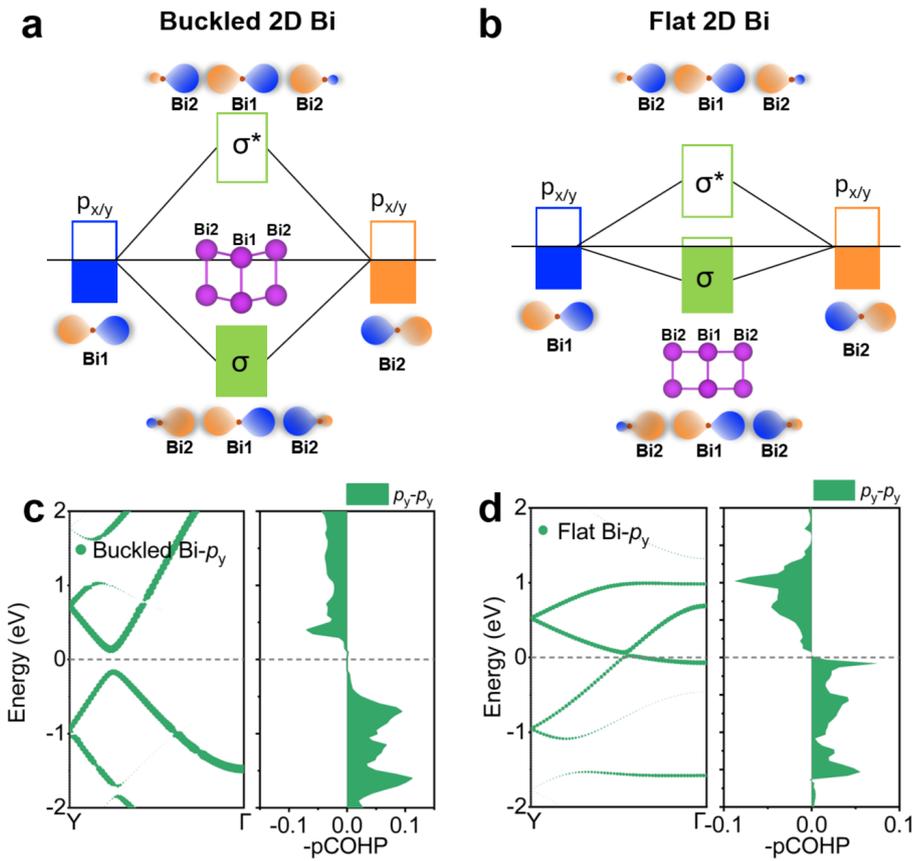

**Figure 2. Orbital interactions of Bi monolayers under different structural configurations.** Schematic illustrations of the formation of bonding (σ) and antibonding (σ*) states between the Bi1-$p_{x/y}$ and Bi2-$p_{x/y}$ in (a) buckled and (b) flat Bi monolayer. Projected band structures contributed by Bi-$p_y$ orbital (left) and projected crystal orbital Hamilton population (pCOHP) curves (right) for the Bi1-$p_y$–Bi1-$p_y$ interactions in the (c) buckled Bi and (d) flat Bi with 17.5% tensile strain. Positive –pCOHP values indicate bonding contributions, while negative values represent antibonding character.

**Contact properties between vdW-squeezed 2D metallic Bi and MoS$_2$**

The contact between bismuth and molybdenum disulfide is of significant interest in current research on 2D transistors.[24,37] The configurations of monolayer Bi in contact with monolayer MoS$_2$ are illustrated in **Figure 3**, and are denoted as Structure A [**Figure 3a**] and Structure B [**Figure 3b**], respectively. These two configurations differ in the choice of lattice constant reference during heterostructure construction: Structure A fixes the MoS$_2$ lattice constant, while Structure B fixes the flat 2D Bi lattice constants. In other words, Structure A involves applying strain to flat 2D Bi to match MoS$_2$, whereas in Structure B, MoS$_2$ is strained to match flat 2D Bi. The corresponding lattice parameters and interlayer distances for both configurations are summarized in **Table S1**.

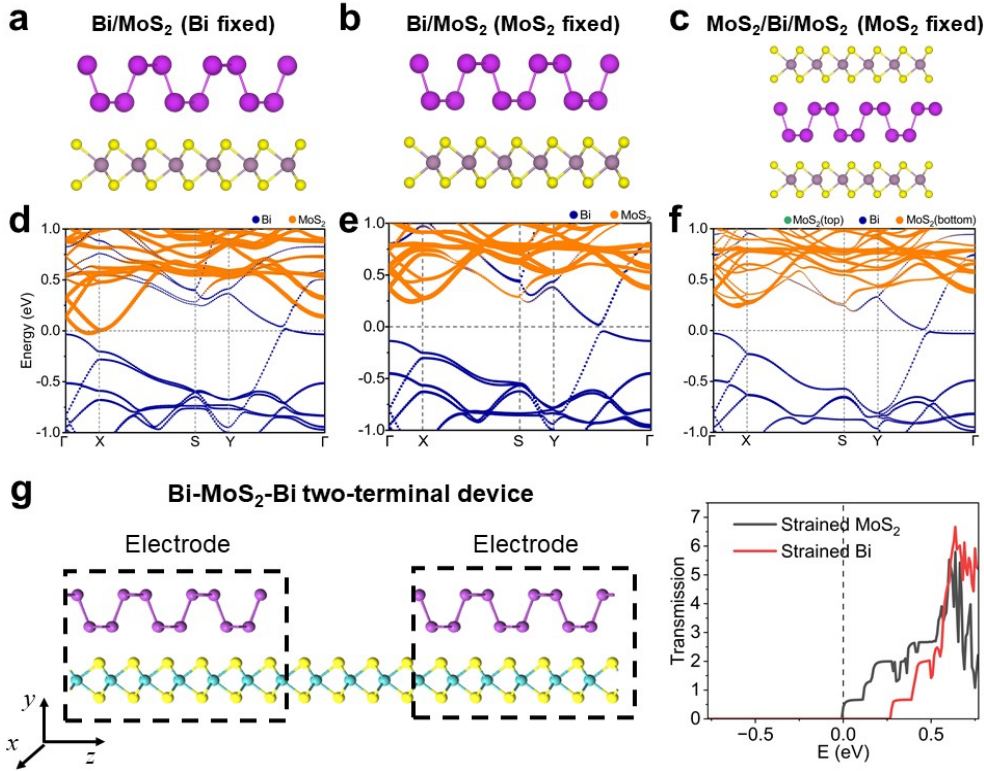

**Figure 3. Contact heterostructure, electronic properties and quantum transport properties of Bi/MoS$_2$.** Geometries of Bi/MoS$_2$ heterostructures constructed by lattice matching to (a) flat 2D Bi and (b) MoS$_2$ lattices, respectively. (c) MoS$_2$–Bi–MoS$_2$ heterostructure constructed by fixing the MoS$_2$ lattice. Band structures corresponding to (a–c), respectively. Orbital contributions from Bi and MoS$_2$ are highlighted. (g) Bi–MoS$_2$ device transport model and (h) the corresponding zero-bias transmission spectrum.

Our calculations reveal that Structure A enables Ohmic contact between Bi and MoS$_2$, while in Structure B, an *n*-type Schottky barrier of 0.243 eV is formed at the interface. We further examine the charge transfer characteristics at the contact regions of both structures (see Supplementary **Figure S2**). The resulting charge transfer distributions show minimal differences between the two configurations, suggesting that variations in interfacial charge transfer do not primarily drive contrasting contact behaviors. According to previous studies, tensile strain applied to MoS$_2$ can reduce its bandgap, causing a downward shift in its conduction band minimum.[38,39] We attribute the emergence of Ohmic contact in Structure A to this strain-induced band modulation, where the tensile strain applied to MoS$_2$ lowers its CBM to align with the work function of flat 2D metallic Bi.

For Structures A and B, we construct a two-terminal device model of Bi–MoS$_2$–Bi [**Figure 3g**] and perform first-principles quantum transport calculations using nonequilibrium Green's function (NEGF) to characterize their zero-bias transport properties.[40,41] The transmission spectra reveal that Structure A exhibits finite transmission at the Fermi level, indicative of Ohmic contact behavior. In contrast, for Structure B, the transmission onset occurs at 0.27 eV above the Fermi level, consistent with an *n*-type Schottky barrier. These transport results are thus in agreement with the electronic band structure DFT calculations.

**Layer-selective Ohmic contact in MoS$_2$-Bi-MoS$_2$ Trilayer Heterostructure**

We further investigated the MoS$_2$–Bi–MoS$_2$ trilayer semiconductor/metal/semiconductor (SMS) heterostructure, denoted as Structure C in **Figure 3c**, which is constructed by fixing the MoS$_2$ lattice. This configuration corresponds to the experimentally fabricated 2D metallic Bi via vdW squeezing method. The calculated band structure of Structure C closely resembles that of Structure B, displaying an *n*-type Schottky contact with a barrier height of 0.243 eV at the Bi– MoS$_2$ interface [**Figure 3f**]. The charge density difference analysis (Supplementary **Figure S2**) reveals a symmetric charge redistribution, with the central flat 2D Bi donating electrons to both the top and bottom MoS$_2$ layers. This symmetric interfacial interaction results in degenerate electronic bands for the two MoS$_2$ layers [**Figure 3f**].

Interestingly, MoS$_2$–Bi–MoS$_2$ SMS heterostructure provides a premise to achieve layer selective charge injection, conceptually illustrated in **Figure 4**. Intrinsically, the bands from the top and bottom MoS$_2$ layers are energetically degenerate [**Figure. 4a**]. The band degeneracy can be lifted by applying an external out-of-plane electric field $E_{field}$. When a sufficiently strong $E_{field}$ is applied, the bands of one MoS$_2$ layer are energetically down-shifted to intersect the Fermi level $E_f$ to form Ohmic contact, while the opposite MoS$_2$ layer remains energetically separated from $E_f$ with a finite Schottky barrier [**Figure 4b**]. The 'layer-location' of the Ohmic and Schottky contacts can be interchanged between the two MoS$_2$ layers by reversing the polarity of $E_{field}$ [**Figure 4c**]. The field-effect tunable band alignment in MoS$_2$–Bi– MoS$_2$ MSM heterostructure thus leads to the concept of *layer-selective Ohmic contact*, representing the *layertronic extension* of conventional Schottky or Ohmic contact across a single MS contact. The layer-selective Ohmic contact enables directional control over the current injection pathways, where a positive $E_{field}$ generates current flow through one $MoS_2$ layer, while a negative $E_{field}$ switches the current flow to the opposite $MoS_2$ layer [**Figure. 4d**].

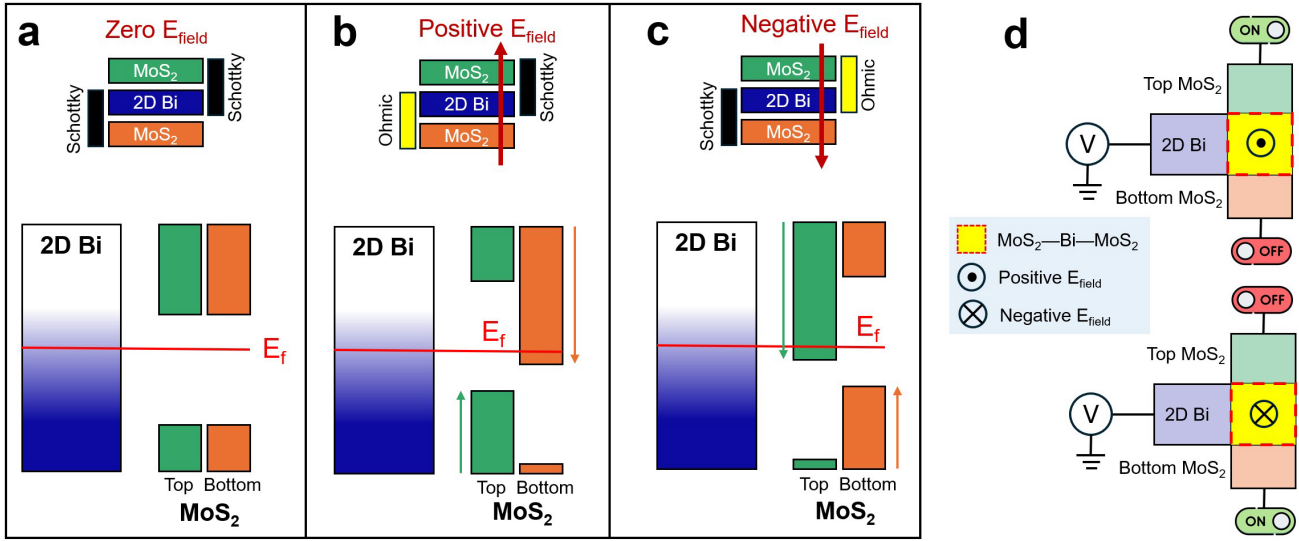

**Figure 4. Concept of layer-selective Ohmic contact in $MoS_2$-Bi-$MoS_2$ trilayer heterostructure.** (a) Shows the band alignment at zero external perpendicular electric field $E_{field}$. The bands of top and bottom $MoS_2$ layers are energetically degenerate. The application of $E_{field}$ breaks the energy degeneracy of bands from top and bottom $MoS_2$ layers. (b) and (c) shows the shifting of the top and bottom $MoS_2$ band structures relative to 2D Bi Fermi level under a positive and negative electric field, respectively. When an appropriately strong $E_{field}$ is applied, one layer forms Ohmic contact while the opposite layer forms Schottky contact, thus giving rise to a *layer-selective* Ohmic contact. Reversing the polarity of $E_{field}$ changes the layer-location of the Ohmic contact. (d) shows the top view of a conceptual layer-selective charge injection device based on $MoS_2$-Bi-$MoS_2$ trilayer heterostructure. Reversing $E_{field}$ spatially switches the current injection pathway from one layer to the other.

To illustrate the concept of layer-selective Ohmic contact, we calculate the band structure evolution of $MoS_2$–Bi–$MoS_2$ under $E_{field}$ [**Figure 5**]. As the electric field increases from 0 to 0.3 V/Å [**Figure 5a-e**], the Initially degeneracy between the two $MoS_2$ layers [**Figure 5a**] is lifted: the band of the top $MoS_2$ layer gradually shifts downward, while that of the bottom $MoS_2$ layer shifts upward, introducing an energy offset between the top and bottom $MoS_2$ layers. When $E_{field}$ reaches 0.15 V/Å, the top $MoS_2$ layer forms an Ohmic contact with Bi, whereas the bottom $MoS_2$ layer still exhibits a Schottky contact [**Figure 5c**]. Upon further increasing the $E_{field}$ beyond 0.15 V/Å, the bands of the bottom $MoS_2$ layer continue to shift upward, eventually leading to Ohmic contact formation between Bi and both $MoS_2$ layers [**Figure 5e**]. Reversing the polarity of the $E_{field}$ leads to an opposite trend [**Figure 5g-k**], i.e. the electronic bands of bottom MoS$_2$ layer energetically *down*-shifted to form Ohmic contact. The MoS$_2$–Bi–MoS$_2$ MSM heterostructure under E$_{field}$ thus offers a platform to realize layer-selective Ohmic contact with potential layertronic device application. Finally, we remark that the E$_{field}$ required to achieve layer-selective Ohmic contact, i.e. E$_{field}$ > 0.15 V/Å (or 15 MV/cm) can be experimentally achieved using high-*k* gate dielectric materials compatible with MoS$_2$, such as the recent experimental demonstration of 5.2-nm-thick GdOCl with breakdown electric field as large as 17.1 MV/cm.[42]

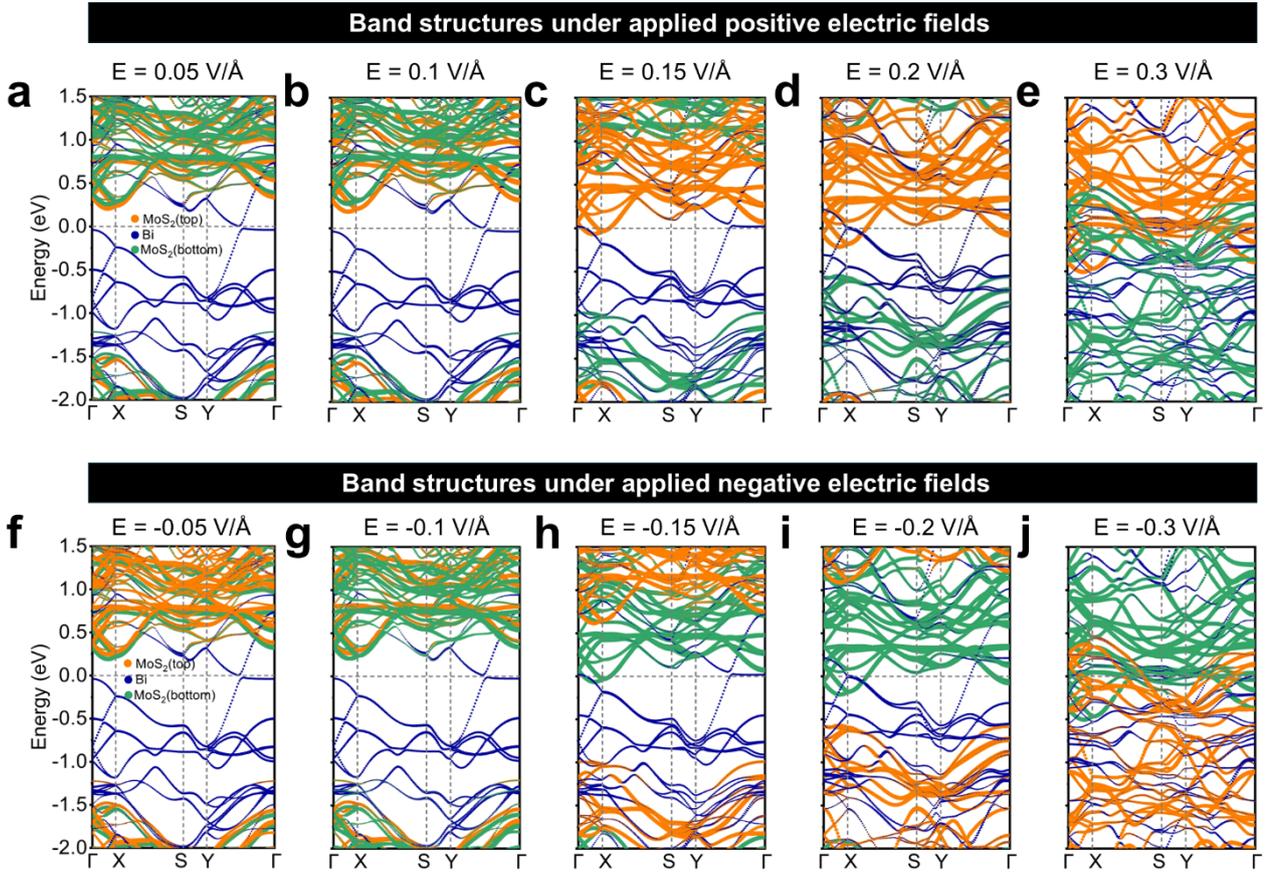

**Figure 5. Electronic band structures of the MoS$_2$–Bi–MoS$_2$ heterostructure under various external electric fields.** (a)–(e) Band structures under positive electric fields correspond to electric fields from 0.05 V/Å to 0.3 V/Å. (g)–(k) Band structures under negative electric fields correspond to electric fields from −0.05 V/Å to −0.3 V/Å.

In conclusion, we proposed that the experimentally observed metallic behavior of monolayer Bi can be explained by pressure-induced flattening of the 2D Bi lattice. We demonstrate that a uniaxial tensile strain of 17.5% flattens the lattice of 2D Bi, and a semiconductor-to-metal phase transition occurs. We further show that the MoS$_2$–Bi–MoS$_2$ trilayer MSM heterostructure provides a premise to explore the concept of layer-selective Ohmic contact in which layer-dependent current injection can be controlled by an out-of-plane electric field to achieve spatial control of current flow in a trilayer MSM heterostructure. These results provide a quantitative explanation of the potential origin metallicity in angstrom-thickness 2D Bi fabricated via vdW-squeezing method, and generalizes one-interface MS Schottky or Ohmic contact into the layertronic case in a two-interface MSM layer-selective Ohmic contact. We propose that the vdW heterostructures composed of other species of ultrathin ångströmthickness 2D metal sandwiched by $MoS_2$ or other transition metal dichalcogenide monolayers to form a fertile ground to unveil the interface physics and device applications of ångström thickness 2D metals.

## Author Contributions

S.W., S.F. and Y.S.A. initiated the ideas. S.W. performed the DFT simulations. Q.L. performed the quantum transport simulations. Y.Y., Z.Y., X.S., J.L., C.S.L., L.K.A., L.-J.L. provided discussion and analysis of results. S.W. and S.F. wrote the manuscript with inputs from all authors. Y.S.A. supervised the project and provided computational resourses.

## Notes

There are no conflicts to declare.

## Acknowledgements

This work is supported by the Singapore National Research Foundation (NRF) under its Frontier Competitive Research Programme (NRF-F-CRP-2024-0001). Q. Li is supported by the high level of research achievements in the cultivation project funding of Hubei Minzu University (No. XN2321), the Educational Commission of Hubei Province of China (No. T201914). Z. Y., X. Y. and J. L. thank the National Natural Science Foundation of China (No. 12274002 and 91964101), the Ministry of Science and Technology of China (No. 2022YFA1203904), High-performance Computing Platform of Peking University, and the MatCloud + high throughput materials simulation engine. T. Sun is supported by the Programs for Science and Technology Development of Henan Province, China (No.252102521070). L. K. A. is supported by the Singapore ASTAR IRG (M23M6c0102).